# Splitting broad beams into arrays of dissipative spatial solitons by material and virtual gratings

Y. J. He,[a,b,c*] B. A. Malomed,[d] F. Ye,[b] J. Dong,[a] Z. Qiu,[a] H. Z. Wang,[a] B. Hu [b,e]

[a] *State Key Laboratory of Optoelectronic Materials and Technologies, Sun Yat-Sen University, Guangzhou, 510275, China*

[b] *Centre for Nonlinear Studies, and The Beijing-Hong Kong-Singapore Joint Centre for Nonlinear and Complex Systems, Hong Kong Baptist University, Kowloon Tong, Hong Kong, China*

[c] *School of Electronics and Information, Guangdong Polytechnic Normal University, 510665 Guangzhou, China*

[d] *Department of Physical Electronics, School of Electrical Engineering, Faculty of Engineering, Tel Aviv University, Tel Aviv 69978, Israel*

[e] *Department of Physics, University of Houston, Houston, Texas 77204-5005, USA*

*Corresponding authors: heyingji8@126.com

**Abstract**

We elaborate two generic methods for producing two-dimensional (2D) spatial soliton arrays (SSAs) in the framework of the cubic-quintic (CQ) complex Ginzburg-Landau (CGL) model. The first approach deals with a broad beam launched into the dissipative nonlinear medium, which is equipped with an imprinted grating of a sufficiently sharp form. The beam splits into a cluster of jets, each subsequently self-trapping into a stable soliton, if the power is sufficient for that. We consider two kinds of sharp gratings – "raised-cosine" (RC) and Kronig-Penney (KP) lattices – and two types of the input beams, fundamental and vortical. By selecting appropriate parameters, this method makes it possible to create various types of solitons arrays, such as solid, annular (with single and double rings), and cross-shaped ones. The second method uses a "virtual lattice", in the form of a periodic transverse phase modulation imprinted into the broad beam, which is passed through an appropriate phase mask and then shone into a uniform nonlinear medium. Two different types of the masks are considered, in the form of a "checkerboard" or "tilings". In that case, broad fundamental and vortical beams may also evolve into stable SSAs, if the beam's power and spacing of the virtual phase lattice are large enough. By means of the latter technique, square-shaped, hexagonal, and quasi-crystalline SSAs can be created.

**I. Introduction**

Spatial solitons have drawn a great deal of interest in optics community and beyond. Recent studies were dealing, in particular, with spatial-soliton arrays (SSAs), which are objects of fundamental importance, and also have a vast potential for applications [1-13]. Various means of generating SSAs in laser cavities [chiefly, in the one-dimensional (1D) form], including the use of external forcing, were investigated [5,6,8,14]. A method for the creation of 2D SSAs in photonic lattices was proposed too [10]. Recently, the stability of SSAs was demonstrated in the framework of the 2D nonlinear Schrödinger (NLS) equation with the cubic-quintic (CQ) nonlinearity and a sharp grating potential, in the form of the Kronig-Penney (KP) lattice, alias a "checkerboard" [15]. The same potential provides for the stabilization of 2D solitons and vortices against supercritical collapse, when the quintic term is self-focusing [16].

*Dissipative* SSAs, including vortical and necklace-shaped patterns [17-22], may be stable in the framework of the complex Ginzburg-Landau (CGL) equation, also with the CQ nonlinearity, as well as in diverse dissipative nonlinear models of cavities driven by external fields [5,6,23-26], in systems described by dissipative Maxwell-Bloch equations [27], and in models based on the saturable noninearity [28,29].

In this work, we present two experimentally relevant methods for creating various types of SSAs in 2D CGL models, including the creation of soliton arrays carrying global vorticity. The first approach assumes shining a broad beam, without or with intrinsic vorticity, into the dissipative nonlinear medium equipped with a grating. If the grating is sharp enough – roughly, being close to the KP lattice – the broad input beam can be efficiently split into to a set of jets, each carrying enough power to self-trap into a stable spatial soliton. We consider the SSA generation by the 2D lattice potentials of two types, *viz.*, the "raised-cosine" (RC) one, and the checkerboard lattice proper. The second approach deals with the uniform nonlinear medium, while a *virtual lattice* is imprinted into the incident beam in the form of a periodic transverse phase modulation. We demonstrate that, by means of both methods, one can construct stable two-dimensional SSAs with various intrinsic structures (including those with embedded vorticity).

It is relevant to mention that a majority of previously published works dealing with 2D solitons in CGL models of laser cavities did not include lattice potentials (gratings), being rather focused on the stability of a single fundamental (zero-vorticity) dissipative soliton in the free space [30]. Recently, the CGL equation with the CQ nonlinearity and square-lattice potential was introduced in Ref. [31], where it was demonstrated that the lattice may efficiently stabilize dissipative *vortex solitons*, built as sets of four local peaks, as well as *crater-shaped* vortices (which are, essentially, squeezed into a single cell of the lattice potential [32]), in the case when the CGL equation that does not include the diffusion term [$\beta = 0$ in Eq. (1), see below], which is relevant to models of large-area laser cavities. Moreover, practically stable three-dimensional (3D) multi-peak complexes carrying the trapped vorticity were found in the 3D model with the 2D lattice potential [33]. However, the creation and control of multi-soliton patterns, i.e., SSAs in 2D dissipative media, by means of gratings has not been considered before, to the best of our knowledge.

The paper is organized as follows. The model based on the material grating is formulated in Section II. Results demonstrating the generation of various types of SSAs in this model by broad input beams, without or with the intrinsic vorticity, are reported in Section III. Then, Section IV presents the second method, based on the virtual phase lattice, and results for the generation of SSAs produced by this method. The paper is concluded by Section V, where we also discuss similarities and differences between the two methods.

**II. The model based on material gratings**

We consider the CQ-CGL equation of the general form [34-36], which is written in terms of laser-cavity models, with coefficients fixed by the usual scaling:

$$iu_z + (1/2)\Delta u + |u|^2 u + v|u|^4 u = iR[u] + V(x,y)u, \qquad (1a)$$

where $\Delta = \partial^2/\partial x^2 + \partial^2/\partial y^2$ is the transverse diffraction operator, $z$ the propagation distance, and the coefficient in front of the cubic self-focusing term is normalized to be 1. Further, $v$ is the quintic self-defocusing coefficient, and the combination of the loss and gain terms is

$$R[u] = \delta u + \beta \Delta u + \varepsilon |u|^2 u + \mu |u|^4 u, \qquad (1b)$$

where $-\delta$ is the linear-loss coefficient, $-\mu$ the quintic-loss parameter, $\varepsilon$ the cubic-gain coefficient, and $\beta$ accounts for the effective diffusion (viscosity). The last term in Eq. (1a) represents the 2D periodic potential induced by the grating. In particular, the RC type of the grating corresponds to

$$V(x,y) = \alpha \left[ \cos(\pi x/d)^m + \cos(\pi y/d)^m \right], \qquad (2)$$

where $d$ and $m$ (which is even) determine the period and sharpness of the lattice, respectively. Obviously, larger $m$ corresponds to a sharper grating. In the limit of $m \to \infty$, the RC goes over into the KP lattice [15,16].

Gratings of a desirable form can be fabricated by means of various technological methods. An especially efficient one is based on writing permanent transverse (2D) structures in bulk materials by means of the laser beams [37].

We consider an input in the form of a broad Gaussian beam launched into the medium,

$$u(r, z = 0) = Ar^S \exp\left(-\frac{r^2}{2w^2}\right) \exp(iS\theta), \quad (3)$$

where $A$ and $w$ represent its amplitude and width, respectively, and integer $S$ is the vorticity (values $S = 0, 1, 2$ are considered below). For simulations, we take $w = 15$, which guarantees that the input beam is broad enough in comparison with the grating's period. The generic case may be adequately represented by fixing the parameters to be

$$\delta = -0.5, \ \beta = 0.5, \ \nu = -0.01, \ \mu = -1, \text{ and } \varepsilon = 2.5. \quad (4)$$

The robustness of the SSAs generated by the beam in direct simulations of Eq. (1a) will be additionally tested by adding noise to them, at the level of 10% of the soliton's amplitude (the additional verification of the stability is relevant, as it is known that, in optical lattice models akin to that considered here, a situation is possible when quasi-stable modes may self-trap from smooth input beams, but eventually they turn out to be unstable against random perturbations [31]). The simulations were performed by means of the standard split-step fast-Fourier-transform method [38].

It is relevant to mention that the viscosity term in Eq. (1b), with $\beta > 0$, is necessary for the stability of all patterns different from the simplest fundamental solitons, such as solitary vortices, in the models of uniform media described by the CGL equation with constant coefficients. However, in the presence of the lattice potential, 2D vortices (and sometimes their 3D counterparts) may be stable too in the model with $\beta = 0$ [31,32,33].

### III. Generation of soliton arrays by material gratings

#### 1. The RC (raised-cosine) lattice

**A. Zero-vorticity input**

Lattice potentials (2) with $m = 6$ and $m = 2$ are plotted in Figs. 1(a) and 1(b). When a broad

Gaussian beam without the intrinsic vorticity ($S=0$) is launched into the dissipative medium with this grating imprinted into it, the beam rapidly splits into a set of jets, when the lattice profile is sharp enough, as shown in Fig. 1(c) for $m=6$. The jets carrying enough power can evolve into stable 2D solitons whose profile is shown in Fig. 2(a), while jets which are too weak quickly disappear, in accordance with the fact that there is a finite excitation threshold in the CQ CGL model [34]. In contrast, if the grating is not sharp enough, the beam does not split, quickly decaying instead, as shown in Fig. 1(d) for $m=2$.

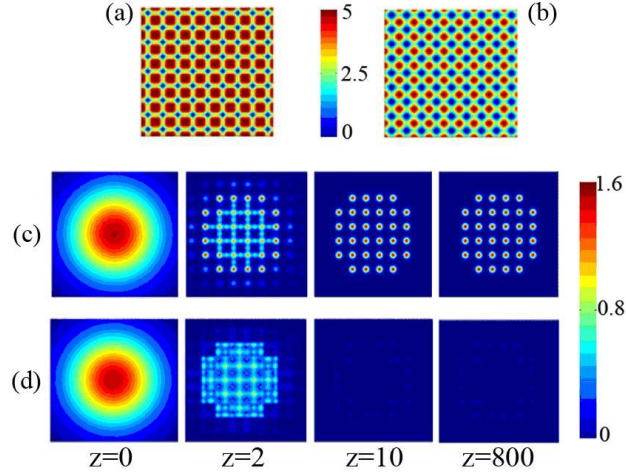

Fig. 1. (Color online) (a) and (b): Profiles of the RC lattice potential (2), with $m=6$ and $m=2$, respectively. The period and depth of the lattices are $d=6.8$ and $\alpha=5$. (c) and (d): The evolution of a broad Gaussian input beam, given by Eq. (3) with $A=1.6$ and $w=15$, under the action of the gratings shown in (a) and (b), respectively. In the panels shown here and in other figures, the coordinate range is $x=y\in(-30,30)$, unless it is specified otherwise. Note that that the characteristic diffraction length is estimated, in the framework of Eq. (1a), as $z_{\text{diffr}}\sim d^2=46.24$. It is seen from this figure and those displayed below that, as a matter of fact, a fraction of this length is sufficient for morphing the stable pattern in (c), and destruction of the unstable one in (d))

It is relevant to stress that, if the formed patterns prove to be stable for the propagation distance shown in Figs. 1 and 2 (and in similar figures displayed below), they actually remains completely stable for an indefinitely long propagation distance (as long as the simulations could be run). In this connection, we note that Eqs. (1b) and (2) give rise to two different natural scales of the propagation distance, *viz*., the diffraction length, $z_{\mathrm{diffr}} \sim d^2$, and the coupling length, which corresponds to the tunneling of light between adjacent cores of the grating structure. The latter length cannot be estimated by a universal formula, as it is inversely proportional to the corresponding tunneling coefficient, which, as is well known, is sensitive to details of the effective potential structure (the comparison of the outcomes of the evolution displayed in Figs. 1(c) and 1(d), as well as many results reported below, clearly illustrate the latter fact). In actual experimental settings, $z_{\mathrm{diffr}}$ may usually be a few millimeters, while the coupling length may vary in broad limits, from $\sim 1$ mm up to several centimeters.

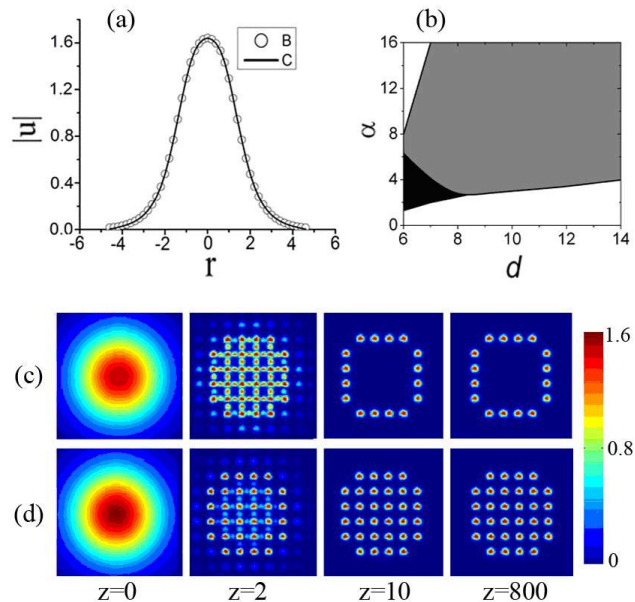

Fig. 2. (Color online) (a) Soliton profiles: a numerical solution (curve B) obtained by means of the imaginary-time integration method, and the result of the self-trapping of one of the jets generated by the splitting of the broad input beam (curve C). (b) In the parametric plane of the lattice depth, $\alpha$, and lattice period, $d$, for fixed $m=6$, solid and annular arrays of the spatial dissipative solitons emerge in the gray and black regions, respectively. Panels (c) and (d) display, severally, the formation of an annular array with $d=7$, $\alpha=3$, and of a solid array with $d=7$, $\alpha=6$.

A typical example of the profile generated by the self-trapping of jets, displayed in Fig. 2(a), confirms that each jet evolves into a stable soliton. Results of the simulations are summarized in Fig. 2(b) which, for fixed $m=6$, shows regions in the parametric plane of the grating's period and strength, $(d,\alpha)$, where solid SSAs and annular SSAs are formed spontaneously (the gray and black region areas, respectively). The annular SSAs emerge only at low values of $\alpha$ and for a sufficiently small period $d$ [e.g., $d<8$ in Fig. 2(b)], because in this case the lattice potential is effectively weak, and cannot prevent the decay of solitons in the central area, making it empty. Examples of the formation of annular and solid SSAs are plotted in Figs. 2(c) and 2(d), respectively.

Next, in Fig. 3(a) we summarize the effect of lattice period $d$ on the formation of SSAs, at different values of $m$. For $6 \leq m \leq 12$, when $d$ is small, annular SSAs are formed [see Fig. 3(b)], while, when $d$ is large enough, solid SSAs emerge, as described above, see Fig. 3(c). The void at the center of the annular SSA becomes larger (if measured in units of $d$) with the further decrease of $d$, as the stronger interaction between the jets causes more solitons to decay in the central area. Next, for $14 \leq m \leq 20$, solid SSAs, cross-shaped ones, and four-soliton complexes successively appear with the decrease of $d$, as shown in Figs. 3(d)-3(f).

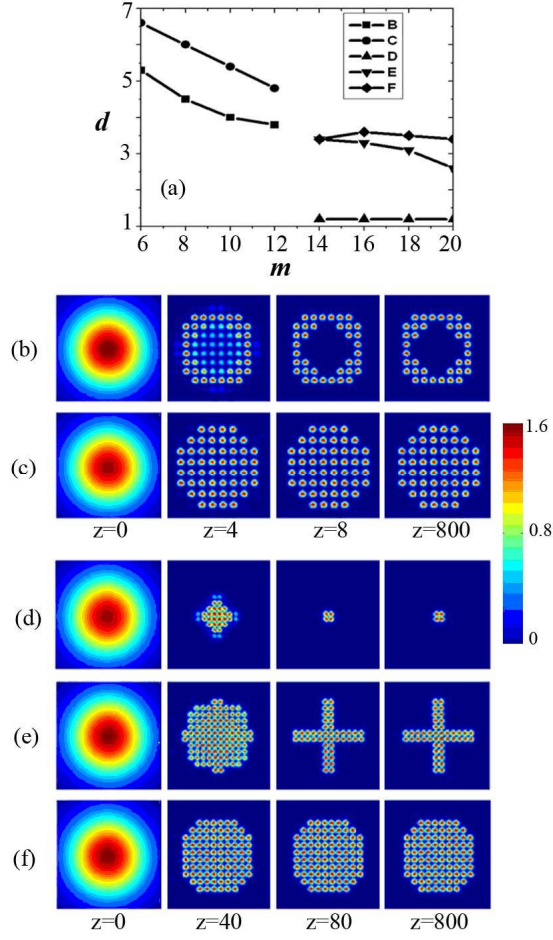

Fig. 3. (Color online) The effect of the variation of lattice period *d* on the shape of SSAs for $\alpha = 4$, at different values of index *m*. (a) For $6 \leq m \leq 12$, annular and solid SSAs emerge, respectively, between curves B and C, and above C, while no arrays are formed below B; for $14 \leq m \leq 20$, four-soliton complexes [see panel (d)] emerge between curves D and E, cross-shaped SSAs – between E and F, and solid SSAs – above curve F, with no SSAs formed below curve D. In panel (a), the lattice period is fixed as *d* =1.2. Other panels display the evolution of the broad Gaussian beam into various stable patterns, *viz.*, an annular SSA in (b), for *m* = 10 and *d* = 5, solid SSA in (c) for *m* = 10 and *d* = 6, a four-soliton complex in (d) for *m* = 18 and *d* = 3, cross-shaped SSA in (e) for *m* = 18 and *d* = 3.4, and a solid SSA in (f) for *m* = 18 and *d* = 4.

Further, we have also considered the effect of amplitude $A$ of the input Gaussian beam on the generation of the SSAs. If the difference of the actual amplitude of the initially generated jets from the amplitude of the soliton solution shown in Fig. 2(a) is too large, the jets cannot evolve towards stable solitons. Respective limits for the formation of SSA are displayed in Fig. 4(a). If $A$ exceeds a critical value, this leads to the disappearances of the jets in the central area, resulting in the generation of hollow SSAs of several types, see Figs. 4(c), 4(d), and 4(f). The void in the annular SSAs becomes larger as $A$ further increases. For $6 \leq m \leq 12$, three kinds of SSAs appear: solid SSAs, single-ring SSAs, and double-ring ones, see Figs. 4(b)-4(d). For $12 \leq m \leq 20$, two types of the patterns are formed: solid SSAs and annular SSAs, at $0.6 \leq A < 2.5$ and $A \geq 2.5$, respectively, as shown in Figs. 4(e) and 4(f).

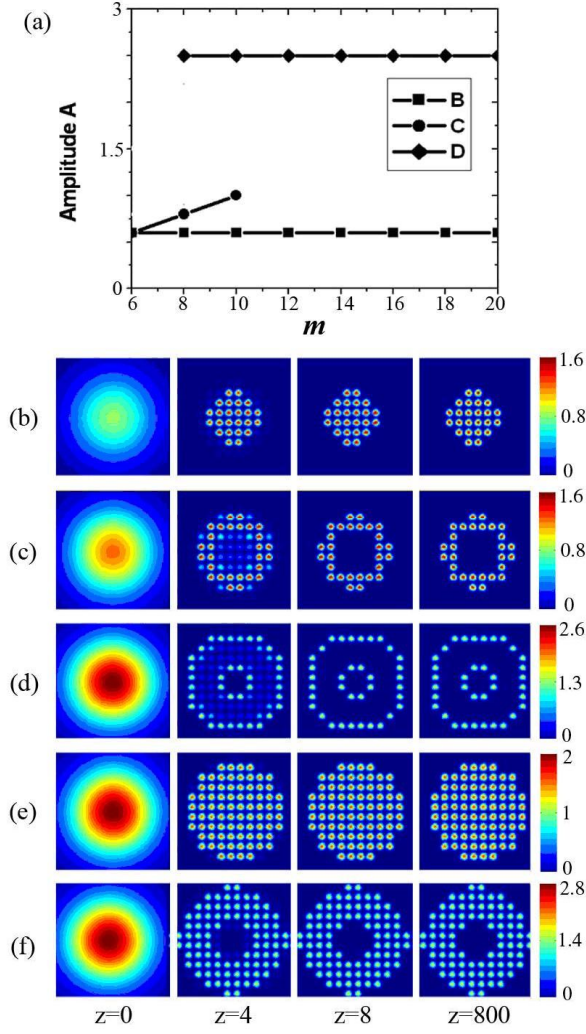

Fig. 4. (Color online) The effect of the variation of amplitude $A$ of the input Gaussian beam (with zero vorticity) on the shape of the SSAs for $\alpha = 4$ and $b = 5$. (a) In the plane of $(m, A)$, no SSAs emerge below line B ($A < 0.6$), at any $m$; for $10 \leq m \leq 20$, annular SSAs appear above line D ($A \geq 2.5$); solid SSAs are formed between lines B and D ($0.6 \leq A < 2.5$). More complex dynamics takes place at $6 \leq m \leq 10$, resulting in the formation of solid SSAs between lines B and C, single-ring SSAs between lines C and D, and double-ring ones above line D ($A \geq 2.5$). Examples of the evolution of the Gaussian input beam: the formation of a stable solid SSA in (b) at $A = 0.8$ and $m = 8$, single-ring SSA in (c) at $A = 1.2$ and $m = 8$, double-ring SSA in (d) at $A = 2.6$ and $m = 8$, solid SSA in (e) at $A = 2$ and $m = 16$, and annular SSA in (f) at $A = 2.8$ and $m = 16$.

**B. Inputs with embedded vorticity ($S \neq 0$)**

The next natural step is to consider effects of vorticity ($S$), embedded into the incident beam, on the formation of SSAs. To this end, Eq. (1a) was simulated with potential (2) and initial condition (3) for $S \neq 0$. Figure 5 shows that the input vortex beam with $S = 1$ may evolve into various patterns, as summarized in panel 5(a). When both the lattice spacing, $d$, and strength, $\alpha$, are large enough, solid SSAs are generated, because in this case the input beam is easily split by the strong grating, see Fig. 5(b). The difference from the solid pattern generated by the input with $S = 0$ is that, as confirmed by the plot of the evolution of the phase field in Fig. 5(c), the pattern *keeps the global vorticity*. When $d$ is smaller, the interaction between the jets strongly affects the established SSA, collecting the power into a single-ring pattern, with the central core formed by four spots, as seen in Fig. 5(d). This structure may be compared to its single- and double-ring counterparts observed in the case of the input with $S = 0$, see Fig. 4(c),(d), the difference being that, in the present case, the structure carries the vorticity, as clearly shown by the phase-evolution images in Fig. 5(e). However, the weak grating, with $\alpha < 2$, cannot split the beam into jets, therefore disordered patterns emerge in lieu of SSAs, see Fig. 5(f).

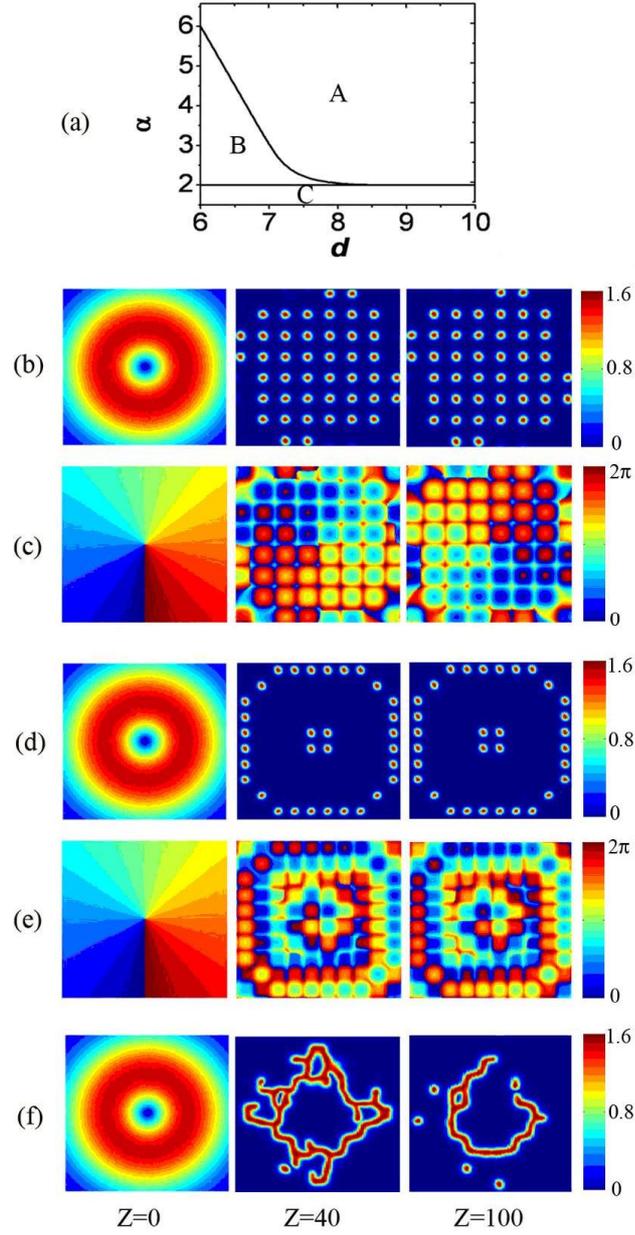

Fig. 5. (Color online) The generation of SSAs by a broad vortex input beam with $S = 1$. (a) The diagram in the plane of ($d, \alpha$) indicates different outcomes of the evolution: region A – solid SSAs which keeps the vorticity, see an example in panels (b,c) for $d = 8$ and $\alpha = 5$; region B – a single-ring annular SSAs with a central core, which also keeps the initial vorticity, see an example in panels (d,e) for $d = 6$ and $\alpha = 5$; region C ($\alpha \leq 2$) – the generation of a disordered pattern, see an example in (f) for $d = 6$ and $\alpha = 0.5$. Panels (b), (d), (f) and (c), (e) display the evolution of the amplitude and phase, respectively. In Figs. 6, 9, and 10 following below, amplitude- and phase-evolution plots are displayed in pairs too.

Figure 6 shows that the input vortical beam with $S = 2$ may also evolve into various patterns, as summarized in panel 6(a). A difference from the case of $S = 1$ is that, instead of the single-ring pattern with the central core [Figs. 5(d),(e)], one here observes either a broad annular structure [Figs. 6(d),(e)], or a double ring, see Figs. 6(f),(g) (the establishment of the latter species of the SSA is observed in a relatively small area C of the parametric plane). Another difference is that, at large values of $\alpha$, the annular structure is a dominating one, rather than the solid pattern, cf. Figs. 5(a) and 6(b). As well as in the case of $S = 1$, all the regular patterns keep the initial vorticity, and a disordered pattern appears at small values of $\alpha$.

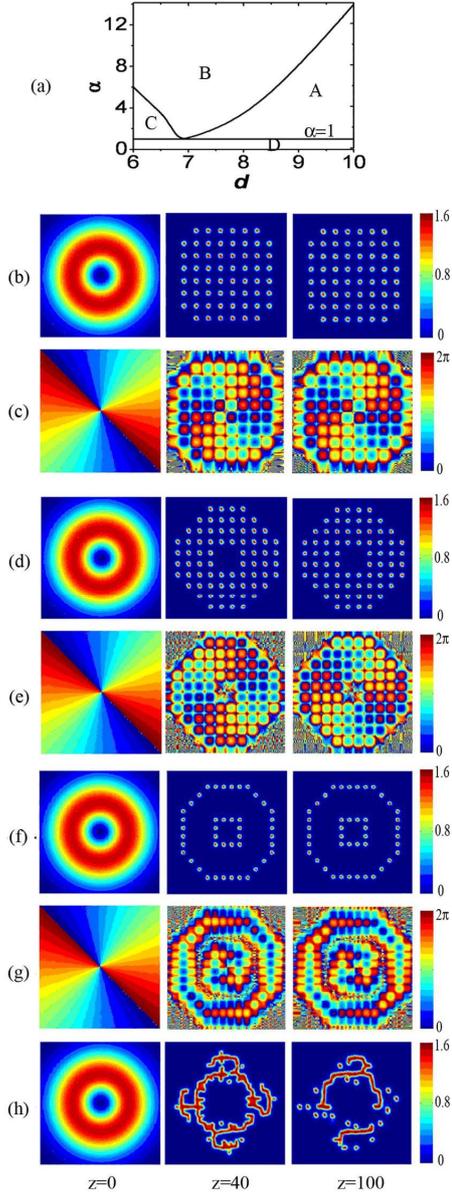

Fig. 6. (Color online) The generation of SSAs by a broad vortical input beam with $S = 2$. In region A of the diagram in panel (a), solid SSAs emerge, see an example in (b,c) for $d = 9$ and $\alpha = 5$. In region B, single-ring annular SSAs appear, see panels (d,e) for $d = 8$ and $\alpha = 10$. In region C, the SSA patterns display a double-ring structure, as shown in (f,g) for $d = 6$ and $\alpha = 5$. All these species of the established patterns keep the initial vorticity. Finally, panel (h) displays an example of a disordered pattern which the input beam generates in region D. The coordinate range shown in this figure is $x = y \in (-45, 45)$.

## 2. The Kronig-Penney lattice

### A. Zero-vorticity input

The KP potential, which was introduced in the context of 2D models in Refs. [15,16], is sharper than RC (2), therefore it is expected to be more efficient as the SSA-sculpting mold. Accordingly, in this case some findings are different from those reported above for the RC potential. Firstly, the minimum lattice period, $d$, which admits the formation of SSAs, is smaller than above ($d_{min} = 4$), where it was $d_{min} = 5$, hence more densely packed SSAs can be generated. In addition, solid SSAs can be created in two intervals of values of the lattice strength, $4 < \alpha < 55$ [see Fig. 7(b)] and $\alpha > 65$ [see Fig. 7(d)]. In the intermediate interval, $55 < \alpha < 65$, the input broad beam can generate only rarefied patterns, as shown in Fig. 7(c).

Effects of the amplitude of the input beam, $A$, on the formation of SSAs were studied too. It was found that annular SSAs are generated when $A$ exceeds a critical value, while solid SSAa emerge at $A < A_{cr}$, see Fig. 8(a). In keeping with this trend, the central void in the annular SSA becomes broader at larger $A$. Examples of the pattern formation in these cases are displayed in Figs. 8(b) and 8(c).

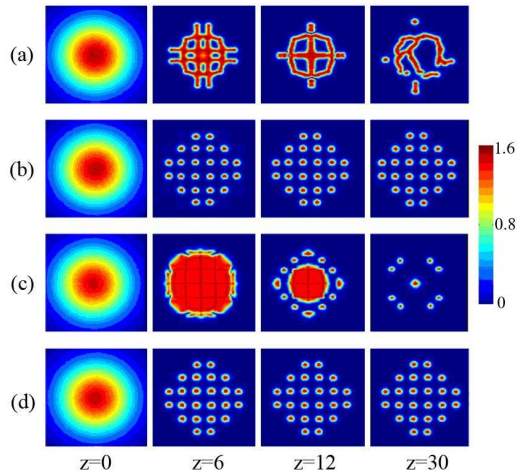

Fig. 7. (Color online) The evolution of broad input beams (with zero vorticity) in the presence of the Kronig-Penney grating with $d$ = 9. (a) No SSA is formed at $\alpha = 3$. (b) The formation of a solid SSA at $\alpha = 10$. (c) A pattern composed by a few solitons survives at $\alpha = 60$. (d) The formation of a solid SSA at $\alpha = 80$.

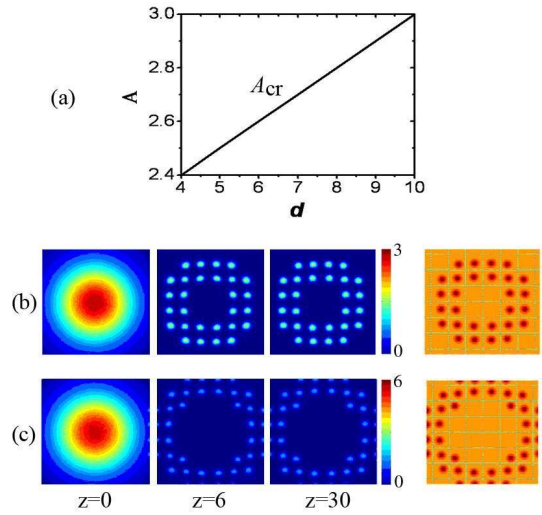

Fig. 8. (Color online) Effects of the variation of amplitude $A$ of the Gaussian input beam (with zero vorticity) on the shape of the SSAs in the KP (Kronig-Penney) lattice. (a) In the plane of lattice period $d$ and amplitude $A$ of the input beam, annular and solid SSAs emerge at $A > A_{cr}$ and $A < A_{cr}$, respectively. The generation of SSAs is shown for $A$ = 3 and $A$ = 6 in panels (b) and (c), respectively. The rightmost panels in rows (b) and (c) display the established SSAs against the background of the KP grating with period $d$ = 9.

**B. Inputs with embedded vorticity ($S \neq 0$)**

Finally, we study the splitting of broad beams, carrying vorticity $S$, into SSAs by the KP grating. With vorticity $S$ = 1 in input (3), various SSAs are generated, as shown in Fig. 9. Solid

SSAs are formed in deep gratings ($\alpha \geq 5$), which are similar to the solid patterns generated by the input with $S = 0$, with the difference that they keep the global vorticity. Disordered patterns, similar to those demonstrated in Figs. 5(f) and 6(h) for $S = 1$ in the model with the RC lattice potential, appear at intermediate values of the strength, $1 < \alpha < 5$. The grating with $\alpha \leq 1$ is so weak that the input beam, maintaining its vortical structure, finally shrinks into a small vortex soliton, squeezed into a square-shaped cluster of four cells of the KP grating, which is surrounded by a satellite set of several fundamental solitons, as shown in Fig. 9(d). The number of the satellites depends on the strength of the grating: for instance, with the decrease of $\alpha$ from 0.5 to 0.2, the number increases from 2 [in Fig. 9(d)] to 7 (not shown here). Examples of the formation of the patterns of different types are displayed in Figs. 9(a),(c),(d), with panels 9(b),(e) showing the respective phase evolution.

Further, Fig. 10 displays the generation of SSAs by broad vortex beams with $S = 2$, shone into the KP grating. Generally, the outcomes of the evolution are similar to those displayed in Fig. 9 for $S = 1$. These include the formation of solid SSAs, see Figs. 10(b),(d) and 10(a),(c), disordered patterns [Fig. 10(f) 10(e)], and the splitting into a complex formed by a compact vortex soliton surrounded by several fundamental ones [Fig. 10(g) 10(f)]. As demonstrated by the respective phase profiles shown in panels 10(c),(e),(h) 10(b),(d),(g), the initial vorticity is kept by all the regular patterns. The difference from the case of $S = 1$ is that the larger vorticity corresponding to $S = 2$ creates a central void, as seen in Figs. 10(b),(d) 10(a),(c).

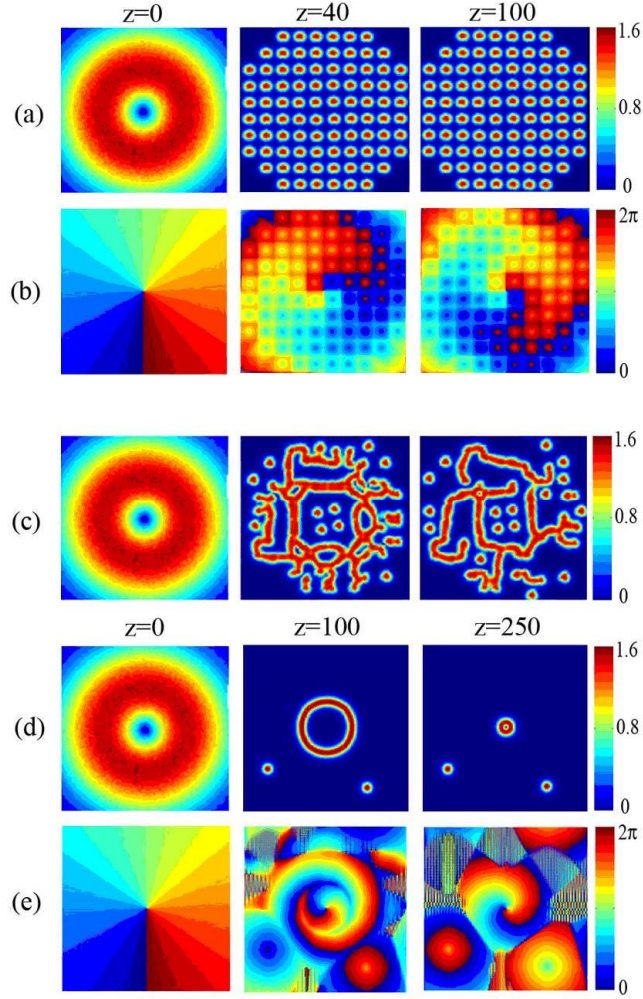

Fig. 9. (Color online) The generation of SSAs by a broad vortical input beam with *S* = 1 coupled into the KP grating. Solid SSAs with the intrinsic global vorticity are formed at $\alpha \geq 5$, as shown in panels (a,b) for *d* = 6 and $\alpha = 10$. Disordered patterns appear at $1 < \alpha < 5$, see an example in panel (c) for *d* = 6 and $\alpha = 3$. For the grating's strength $\alpha \leq 1$, the beam evolves into a compact vortex soliton and several fundamental ones, as shown in panels (d,e) for *d* = 6 and $\alpha = 0.5$.

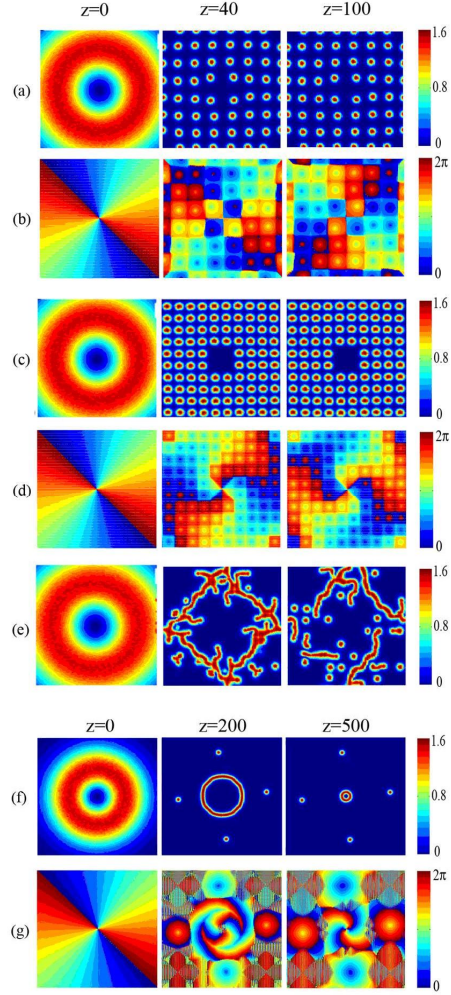

Fig. 10. (Color online) Typical examples of the generation of SSAs by a broad vortical input beam with $S = 2$ coupled into the KP grating. Solid SSAs with the inner void, which is supported by the trapped vorticity, are formed at $\alpha \geq 5$, see respective examples in panels (a,b) for $d = 9$ and $\alpha = 10$, and (c,d) for $d = 6$ and $\alpha = 10$ (the difference between them is that the one corresponding to the larger values of the grating's spacing, $d$, is less stiff, hence it is stronger deformed by the central void). The disordered pattern appears at $1 < \alpha < 5$, see an example in panel (e) for $\alpha = 3$. For values of the grating's strength $\alpha \leq 1$, the beam evolves into a vortex soliton, which, at $0.2 \leq \alpha \leq 0.8$, is accompanied by a satellite set of four fundamental solitons, as shown in panels (f,g) for $d = 6$ and $\alpha = 0.5$. The coordinate ranges are $x = y \in (-30, 30)$ in panels in (a)-(e), and $x = y \in (-45, 45)$ in (f) and (g).

**IV. Generation of soliton arrays by the phase modulation**

In this section, we address the CGL model of the uniform nonlinear medium, based on Eq. (1a) with $V(x,y) = 0$. However, the input is taken as the Gaussian beam (possibly, a vorticity-carrying one), passed through a phase mask which imparts a periodic phase modulation, $\Delta\varphi = f(x,y)$, to the beam:

$$u(r, z=0) = Ar^S \exp\left(-\frac{r^2}{2w^2}\right)\exp[iS\theta + if(x,y)]. \tag{5}$$

**A. Zero-vorticity input**

Figures 11(a) and 11(f) show two typical examples of the evolution of the modulated Gaussian beams into stable SSAs, provided, respectively, by the use of *checkerboard* phase masks with he phase difference of π between adjacent cells [see Fig. 11(b)], and *tile* masks, formed by in-phase cells separated by narrow troughs phase-shifted by π, see Fig. 11(d). In fact, we considered more general cases too, with the phase jumps differing from π. When the beam is passed through the phase mask, the phase jumps slice it into an array of jets, with the adjacent out-of-phase and in-phase jets interacting repulsively and attractively, respectively [39]. With the parameters fixed as per Eq. (4), it was found from the simulations that only the jets possessing power $E \geq 17$ could evolve into stable fundamental solitons, whose profile as shown in Fig. 12(a). At $E < 17$, the jets would decay.

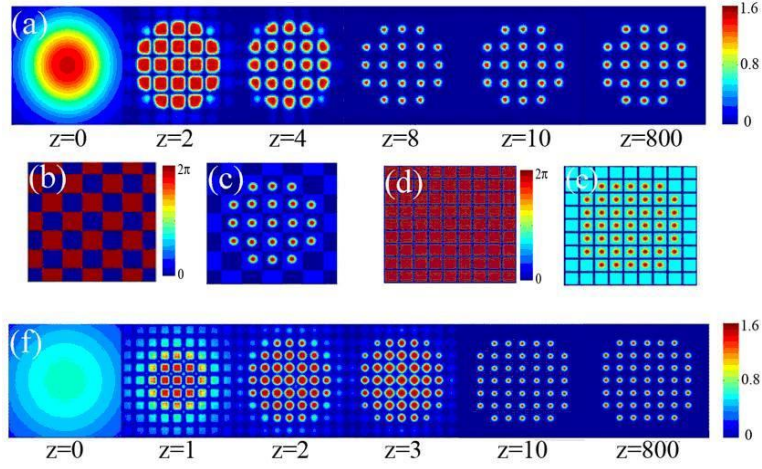

Fig. 11. (Color online) Square-lattice spatial soliton arrays formed by phase-modulated Gaussian beams. (a) The transformation of the beams, with $A = 1.6$ and $w_{x,y} = 16$, into stable arrays induced by the modulation imposed by the checkerboard mask, which is shown in panel (b). (c) The established array shown against the background of the checkerboard mask. (d) The tiling mask. (e) The established array shown against the background of the tiling mask. (f) The transformation of the modulated Gaussian beam, with $A = 0.7$ and $w_{x,y} = 44$, into the stable array in the case of the tiling mask. The coordinate range is $x = y \in (-30, 30)$ in panels (a)-(c), and $x = y \in (-45, 45)$ in (d) and (f).

The period of the phase lattice should be large enough, otherwise interactions between solitons in the emerging array will destroy it (quite similar to the conclusion made above for material gratings). The necessary minimum distance between adjacent solitons can be predicted by using balance equations for the energy and momentum [40], along with the appropriate perturbation theory [39,41,42]. To this end, we use the evolution equations for the total power, $E[u]$, of the conservative part of Eq. (1a):

$$\frac{dE[u]}{dz} = \int_{-\infty}^{\infty}\int_{-\infty}^{\infty}[u^*(iR[u]) + u(iR[u])^*]dxdy$$

$$= 2\int_{-\infty}^{\infty}\int_{-\infty}^{\infty}[\alpha|u|^2 + \varepsilon|u|^4 + \mu|u|^6 - \beta(|u_x|^2 + |u_y|^2)]dxdy \equiv F[u], \quad (6)$$

where the asterisk stands for the complex conjugation. To apply the perturbation theory, we consider a pair of far separated solitons with a certain phase shift between them, $\sigma = 0$ or $\pi$:

$$u(r) = u_0\left(\left|\overline{r} - \overline{d}/2\right|\right)\exp(i\sigma) + u_0\left(\left|\overline{r} + \overline{d}/2\right|\right), \quad (7)$$

where $u_0(r)$ is assumed to be a stable soliton solution of Eq. (1a) (in the absence of the lattice potential, $V = 0$), $\overline{r} \equiv \{x, y\}$, $r = \sqrt{x^2 + y^2}$, and $\overline{d}$ is the vector accounting for the separation between the solitons. For stationary solutions, the power is constant, hence the corresponding solution must represent fixed points of Eq. (6), $F[u_0,d] = 0$, as predicted by the substitution of ansatz (7) into Eq. (6). Figure 12(b) shows that, both for $\sigma = \pi$ and $\sigma = 0$, the power-change rates – $F_1$ and $F_2$, respectively – drop practically to zero at $d > d_{min} \approx 8.8$, hence stationary patterns may be formed by practically non-interacting solitons at such sufficiently large values of the separation between them.

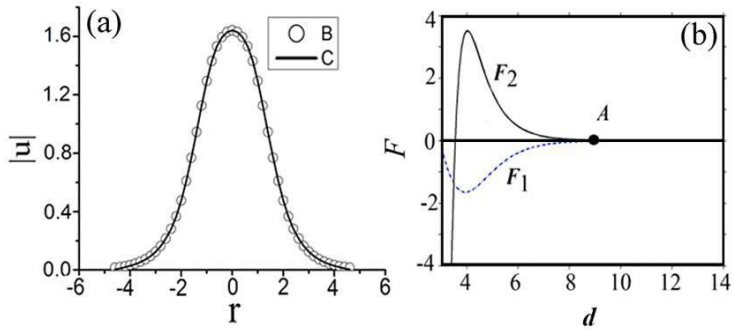

Fig. 12. (Color online) (a) The profile of a stable 2D dissipative soliton, as obtained from a direct numerical solution (curve B), and the result of the evolution of a jet originating from the splitting input beam (curve C). (b) The power-evolution rates, $F_1$ and $F_2$, versus the separation between the two solitons, $d$, which form pair (7),

for the phase shift $\sigma = \pi$ and $\sigma = 0$, respectively. The figure indicates that both $F_1$ and $F_2$ vanish at $d > d_{\min} \approx 8.8$ (to the right of point A), for the parameters chosen as per Eq. (4). A formal fixed point, $F_2(d \approx 3.5) = 0$, is irrelevant, as at this value of $d$ the two solitons strongly overlap, making the underlying ansatz (7) irrelevant.

Next, we consider a hexagonal phase mask built of circular tiles, see Fig. 13(a), which naturally generates a hexagonal array [Fig. 13(b)]. In addition, the hexagonal mask can be created with a defect [Fig. 13(c)], which gives rise to a vacancy in the SSA, see Fig. 13(d). For the formation of stable hexagonal SSAs, the ratio of the area occupied by the tiles to that of troughs separating them, which is denoted $k\%$, should be neither too small nor too large, as otherwise the effective contrast of the phase-modulation pattern, imprinted by the phase mask, is not strong enough to generate a stable array. The appropriate values of $k\%$ belong to parameter regions shown in Figs. 13(e) and (f).

Using this method, SSAs with any desired structure can be readily produced, as the corresponding phase mask can be easily designed by means of the numerical computation, and then fabricated by means of the reactive-ion etching technique. An interesting example is an *N*-fold quasi-crystalline structure. Quasicrystals have drawn much attention due to their peculiar properties, such as broad absolute photonic bandgaps [43,44], but nonlinear quasi-crystalline patterns are difficult to produce. Here, we present an example of the generation of a stable *dodecagonal* (12-fold) quasi-crystalline SSA, which has not been reported before, as far as we know. The respective phase masks are shown in Figs. 14(a) and (d). Under the action of the modulation imposed by such masks, the broad beams indeed evolve into dodecagonal SSAs featuring two different shapes, as demonstrated in Figs. 14(c) and 14(f).

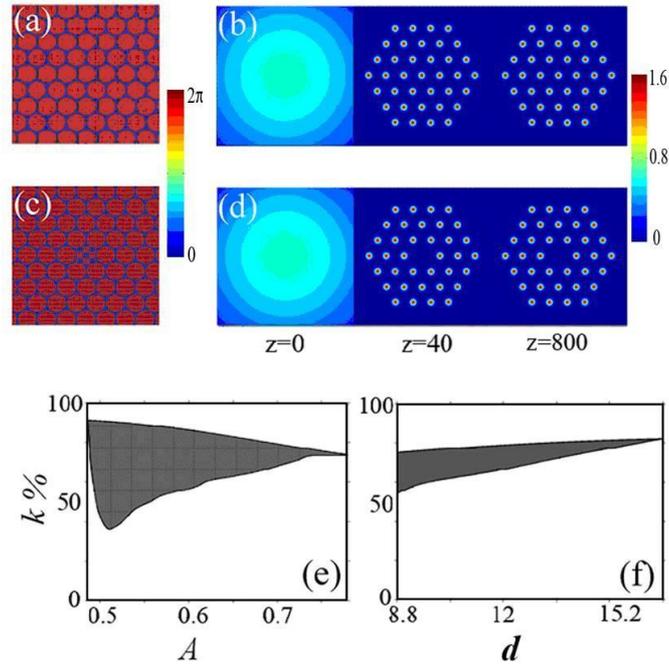

Fig. 13. (Color online) Soliton arrays created by hexagonal phase masks. (a) The mask without the defect. (b) The transformation of the modulated Gaussian beam into a defect-free stable array. (c) The phase mask with a defect at the center. (d) The transformation of the input Gaussian beam, phase-modulated by the mask, into a stable array with the defect. Parameters of input beam (3) (with $S = 0$) are: A = 0.7, wx,y = 44, and the modulation period is d = 12. (e) The relation between k% and A for d = 12 (see the text). (f) The relation between k% and d for A = 0.7. Soliton arrays are generated in the shaded areas in (e) and (f). The coordinate range in panels (a)-(d) is $x = y \in (-45, 45)$.

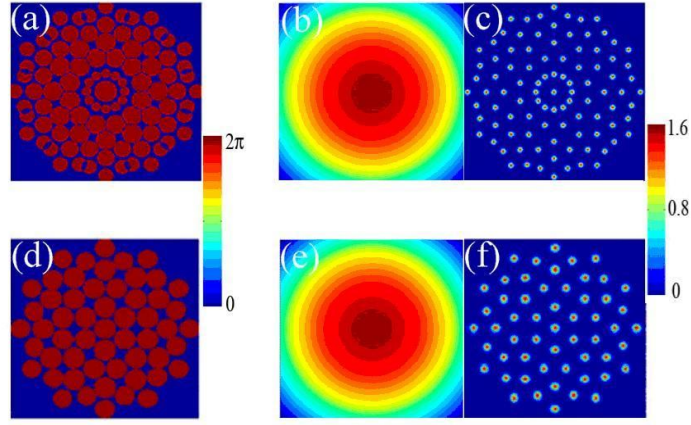

Fig. 14. (Color online) Two species of dodecagonal (12-fold) quasi-crystalline spatial-soliton arrays. (a,d): The respective phase masks. (b,e): The incident Gaussian beam with $A = 0.55$, and $w_{x,y} = 140$ in (b), and 90 in (e). The generated stable dodecagonal arrays are displayed in panels (c) and (f). The coordinate range is $x = y \in (-75, 75)$ in panels (a-c), and $x = y \in (-50, 50)$ in (d-f).

**B. Inputs with the vorticity ($S \neq 0$)**

The evolution of a vortical beam with the lowest embedded vorticity, $S = 1$, into SSAs under the action of the phase modulation, imposed by the same checkerboard mask as in Fig. 11(b), is displayed in Fig. 15. For $d \geq 10$, the vortical beam evolves into a SSA with a compact vortex trapped at the center. For $5 < d < 10$, the same input evolves into a SSA with a void at the center, which may be understood as a result of destruction of the central vortex by the strong interaction with surrounding jets. Under the action of the phase-modulation pattern with a smaller spacing, $d \leq 5$, the input beam generates a disordered pattern, due to still stronger interaction effects.

Figure 16 demonstrates that a vortical input beam with $S = 2$ evolves into SSAs following different scenarios. For $d \geq 15$, it generates an asymmetric array. For $5 < d < 15$, the vortical beam evolves into a SSA with the inner void, and at $d \leq 5$, the beam creates a disordered pattern. The two latter outcomes of the evolution are similar to the situation in the case of $S=1$, cf. Fig. 15.

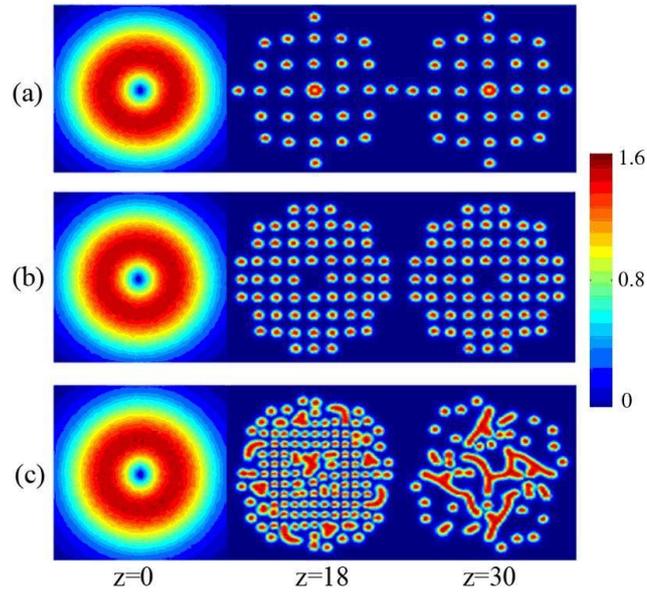

Fig. 15. (Color online) The generation of soliton arrays by imprinting the checkerboard phase modulation onto a broad incident vortical beam with $S$ = 1. (a) $d$ = 11.6, (b) $d$ = 7.8, and (c) $d$ = 4.2. The coordinate range is $x = y \in (-39, 39)$.

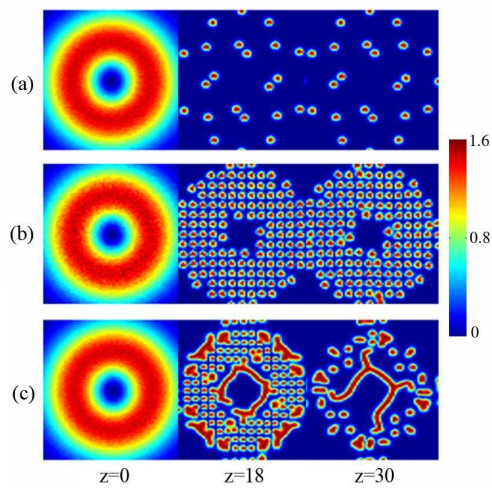

Fig. 16. (Color online) The generation of soliton arrays by the application of the checkerboard phase modulation to a broad incident vortical beam with $S = 2$. (a) $d = 18.6$, (b) $d = 9.3$, and (c) $d = 4.2$. The coordinate range is $x = y \in (-39, 39)$.

## V. Conclusions

In this work, we have proposed and theoretically elaborated two methods for the creation of SSAs (spatial soliton arrays) of diverse types in dissipative media, described by 2D models of the CQ-CGL type. The first method is based on coupling a broad Gaussian input beam, with or without intrinsic vorticity, into the medium with an imprinted periodic grating. The input beam is split by the grating into a cluster of jets, provided that the corresponding periodic potential is sharp enough, being represented by the RC (raised-cosine) or KP (Kronig-Penney) lattice. The jets rapidly self-trap into stationary SSAs, if they carry enough power. Adjusting the intensity and vorticity of the input beam, and the structure and period of the grating, one can create SSAs of various types, such as solid and single- or multiple-ring-shaped ones, as well as crosses and four-soliton complexes. The second method relies on passing the broad beam through a phase mask before injecting it into the uniform medium. The phase modulation imposed by the mask (which can be designed, e.g., in the form of a "checkerboard" or hexagonal "tiling") also induces splitting of the beam into clusters of jets, which can then self-trap into soliton arrays, if the power is sufficient, and the mask spacing is large enough. The latter method also makes it possible to design SSAs with various structures, including squares, hexagons, and quasi-crystals.

An essential difference between the two methods is that the SSAs generated by the material grating are genuine multi-soliton bound states, in which the interaction between adjacent solitons is balanced by their pinning to the underlying lattice potential. On the other hand, the arrays generated in the uniform medium by the initial phase modulation are effectively stable patterns because, in the case of a sufficiently large spacing, the interaction forces are two weak to deform the arrays. In the latter case, the patterns are essentially stabilized by the effective viscosity, accounted for by the term $\sim \beta$ in Eq. (1b). It gives rise to a friction force [45], which prevents the motion of the solitons under the action of the residual weak interaction forces. In the presence of the material grating, the viscosity is not necessary [31]. It is relevant to stress that the creation of multi-soliton arrays by means of periodic phase modulations imprinted onto the broad input beam was not considered in previous studies of models with the CQ nonlinearity.

The results reported in this work suggest new experiments aimed at testing the physics of multi-soliton complexes. The findings may also be potentially useful for the design of all-optical multi-channel data-processing schemes, as well as for engineering dynamical multi-pixel patterns (cf. Ref. [29]). A challenging direction for further studies in this field is to consider similar possibilities in 3D models.

**Acknowledgments**

This work was supported by the Hong Kong Baptist University and the Hong Kong Research Grants Council, Guangdong Province Natural Science Foundation of China (Grant No. 9451063301003516).